\documentclass[a4paper]{jpconf}
\usepackage{graphicx}

\usepackage{amsbsy,amssymb,latexsym,amsfonts,amsmath}
\usepackage{graphics,color}

\allowdisplaybreaks

\newcommand{\cG}{{\mathcal G}}

\newcommand{\cL}{{\mathcal L}}
\newcommand{\cF}{{\mathcal F}}
\newcommand{\cR}{{\mathcal R}}

\newcommand{\al}{\alpha}
\newcommand{\be}{\beta}
\newcommand{\ga}{\gamma}
\newcommand{\de}{\delta}
\newcommand{\ep}{\epsilon}

\newcommand{\la}{\lambda}

\newcommand{\str}{{\rm Str}}
\newcommand{\alg}[1]{\mathfrak{#1}}
\newcommand{\el}{\nonumber}

\newcommand{\pexp}{{\rm Pexp}}

\begin{document}

\begin{flushright}
{KUNS-2520} 
\end{flushright}

\title{Integrable deformations of the AdS$_5\times$S$^5$ superstring and the classical Yang-Baxter equation \\
{\it -- Towards the gravity/CYBE correspondence -- }}  

\author{Takuya Matsumoto}
\address{Nagoya University, Department of Mathematics, Nagoya 464-8602, Japan.}  
\ead{takuya.matsumoto@math.nagoya-u.ac.jp} 
\vspace{2mm}
\address{Institute for Theoretical Physics and Spinoza Institute, Utrecht University, \\
Leuvenlaan 4, 3854 CE Utrecht, The Netherlands.} 

\author{Kentaroh Yoshida}
\address{Department of Physics, Kyoto University, Kyoto 606-8502, Japan.} 
\ead{kyoshida@gauge.scphys.kyoto-u.ac.jp}

\begin{abstract}
Based on the formulation of Yang-Baxter sigma models developed by Klimcik and Delduc-Magro-Vicedo, 
we explain that various deformations of type IIB superstring on AdS$_5\times$S$^5$ 
can be characterized by classical $r$-matrices satisfying the classical Yang-Baxter equation (CYBE). 
The relation may be referred to as
{\it the gravity/CYBE correspondence}. We present non-trivial examples of the correspondence
including Lunin-Maldacena backgrounds for $\beta$-deformations of the ${\cal N} = 4$ super
Yang-Mills theory and the gravity duals for non-commutative gauge theories. 
We also discuss non-integrable backgrounds such as AdS$_5\times T^{1,1}$ as a generalization. 
\end{abstract}

\section{Introduction}

The AdS/CFT correspondence between type IIB superstring 
on the AdS$_5\times$S$^5$ background and the ${\cal N}=4$ super Yang-Mills theory \cite{M}  
has a rich integrable structure. 
On the string-theory side, the Green-Schwarz action on AdS$_5\times$S$^5$ 
is described as a sigma model on a supercoset $PSU(2,2|4)/[SO(4,1)\times SO(5)]$ \cite{MT}. 
Based on the ${\mathbb Z}_4$-grading of the superconformal group $PSU(2,2|4)$\,, 
a novel one-parameter flat connection has been constructed \cite{BPR} 
(For an argument in the Roiban-Siegel formulation \cite{RS} see \cite{Hatsuda}).  
On the gauge-theory side, one-loop dilation operator can be mapped to 
the hamiltonian of an integrable spin chain \cite{MZ}. For a comprehensive review, see \cite{review}.  
However, the fundamental mechanism underlying the duality is still unclear. 
To understand the duality much deeper, 
it would be of significance to argue its deformations which preserve the integrability. 
The deformations allow us to take various limits by adjusting deformation 
parameters and make our understanding much deeper. 

A novel prescription to deform principal chiral models was proposed by Klimcik, 
which is the so-called {\it the Yang-Baxter sigma models} \cite{Klimcik}.  
The key ingredient in this prescription is {\it classical $r$-matrices} 
satisfying the {\it modified classical Yang-Baxter equation (mCYBE)}. 
Here it is assumed that the $r$-matrices are constant and do not depend on  
spectral parameters. If the $r$-matrices are skew-symmetric,    
the integrability of the sigma models is preserved. 
Then this prescription was generalized to symmetric coset cases by Delduc-Magro-Vicedo \cite{DMV}. 
Just after that, they succeeded to construct a $q$-deformed action of the AdS$_5\times$S$^5$ superstring \cite{DMV2}. 
Although in the original formulation, the mCYBE was utilized, 
the usual CYBE is also available to describe integrable deformations along a similar line \cite{KMY-Jor}.  
Thus one may consider two kinds of integrable deformations of the AdS$_5\times$S$^5$ superstring\footnote{%
For another approach to integrable deformations, see \cite{HMS1,S}. 
};
\begin{enumerate} 
\item integrable deformations from the mCYBE,  
\item integrable deformations from the CYBE. 
\end{enumerate}
As we will explain later, the difference between them  should not be confused.

In the case (i), a typical skew-symmetric constant $r$-matrix 
is of the Drinfeld-Jimbo type \cite{Drinfeld1,Jimbo}. 
The deformed string action \cite{DMV2} exhibits the quantum group symmetry  
${\cal U}_q(\alg{psu}(2,2|4))$ \cite{DMV3}\footnote{For earlier arguments on a $q$-deformed $\mathfrak{su}(2)$ 
in squashed sigma models, see \cite{KYhybrid,KMY-QAA,KOY}.}. 
Thus, this type of deformation might be regarded as a $q$-deformation of 
the AdS$_5\times$S$^5$ string action.  
The metric and NS-NS two-form of the deformed background has been explicitly obtained by \cite{ABF}. 
However, completing the full supergravity solutions is still open problem. 
There are several related works in this direction. 
Some specific limits of the deformation parameter are investigated in \cite{HRT} and \cite{AvT,AdLvT}. 
Especially, the relation to the AdS$_5\times$S$^5$ mirror models was found in \cite{AvT,AdLvT}.
A fast-moving string limit and its relation to anisotropic Landau-Lifshitz sigma models have been 
studied by \cite{Kame}.  
A deformed Neumann model was presented by \cite{AM} with a rotating string ansatz 
on the deformed AdS$_5\times$S$^5$\,.
Classical string solutions, called giant magnons, 
on the deformed background are constructed in \cite{AdLvT,Magnon}. 
A new coordinate system is proposed in \cite{Kame2} and minimal surface solutions 
have been discussed. Interestingly, it exhibits a BPS-like behavior, while supersymmetries 
are $q$-deformed and these are not realized in the usual sense any more. 

In the case (ii), on the other hand, 
an application for the AdS$_5\times$S$^5$ superstring is proposed in \cite{KMY-Jor}. 
This case contains Jordanian type deformations \cite{Jordanian,KLM} and 
often called Jordanian deformations of the AdS$_5\times$S$^5$ superstring. 
This short article is devoted to give a review of integrable deformations of this type.  
As a remarkable feature of the deformations based on the CYBE, 
there are several types of $r$-matrices such as  
{\it abelian, Jordanian} and {\it abelian-Jordanian} types. 
In addition to varieties of the $r$-matrices, 
one can consider partial deformations of the AdS$_5\times$S$^5$ background. 

An important question is what the physical meaning of the CYBE-type deformation is. 
Our conjecture for this question is that there exists a duality chain which gives the 
identical supergravity background defined by the CYBE-type deformation. 
In other words, we might interpret the solutions of the CYBE as a moduli space of the 
deformations of the AdS$_5\times$S$^5$ string background.  
In fact, based on a series of our works \cite{MY-LM,MY-MR,KMY-SUGRA,MY-TsT,CMY-T11},
we will present some non-trivial examples including the Lunin-Maldacena 
backgrounds \cite{LM,Frolov} and the gravity duals of the non-commutative gauge theories \cite{HI,MR}. 
More generally, we expect that there exists a correspondence between the classical $r$-matrices satisfying 
the CYBE and the deformed type IIB supergravity backgrounds. 
We refer to this relation as {\it the gravity/CYBE correspondence} \cite{MY-LM}. 

This article is organized as follows. 
After recalling the CYBE and classical $r$-matrix in Section \ref{Sec:classical-R}, 
we introduce a deformed AdS$_5\times$S$^5$ superstring action in Section \ref{Sec:action}. 
In Section \ref{Sec:twist}, as a generic property of the CYBE-type deformation, 
we will show that CYBE-type deformations may be interpreted as non-local gauge transformations 
of the left-invariant one-form. 
In Section \ref{Sec:CYBEtograv}, we explicitly derive the deformed metric and 
the NS-NS two-form, and present some concrete examples of the 
gravity/CYBE correspondence.  
We also demonstrate that our deformation technique works beyond the integrability. 
As an example, in Section \ref{Sec:T11}, 
we consider deformations of the AdS$_5\times T^{1,1}$ background.  
Section \ref{Sec:concl} is devoted to conclusion and discussion.

\section{Classical $r$-matrix \label{Sec:classical-R} } 

In this section, we will give a short review on some properties of the CYBE 
and classical $r$-matrices as solutions of the CYBE.  

\subsection{Classical Yang-Baxter equation (CYBE)} 

Let $\alg{g}$ be a Lie algebra. A {\it classical $r$-matrix} (in tensorial notation) 
\[
r=\sum_i a_i\otimes b_i \quad\in\quad \alg{g}\otimes \alg{g}
\] 
is defined as a solution of {\it the classical Yang-Baxter equation (CYBE)}; 
\begin{align}
[r_{12},r_{23}]+[r_{12},r_{13}]+[r_{13},r_{23}]=0\,,
\label{CYBE-ten}
\end{align}
where $r_{12}=r\otimes 1, r_{23}=1\otimes r$ and $r_{13}=\sum_i a_i\otimes 1 \otimes b_i$
with a schematic notation. 
The CYBE is regarded as a classical analogue of the quantum Yang-Baxter equation (YBE);
\begin{align}
\cR_{12}\cR_{13}\cR_{23}=\cR_{23}\cR_{13}\cR_{12}\,. 
\end{align} 
Indeed, plugging an expansion \[
\cR_{ij}=1+\hbar r_{ij}+{\cal O}(\hbar^2)
\] 
with the YBE, it is easy to derive the CYBE \eqref{CYBE-ten} at ${\cal O}(\hbar^2)$\,. 

When there exists a non-degenerate symmetric bilinear two-from 
$\langle~,~ \rangle:\alg{g}\otimes\alg{g}\to\mathbb{C}$\,, one can introduce 
a linear operator denoted by $R\in {\rm End}(\alg{g})$ by 
\begin{align}
R(x) = \langle r,x\rangle_2=\sum_i a_i \langle b_i,x\rangle 
\qquad \text{for}\qquad {}^\forall x\in \alg{g}\,. 
\label{linear-R}
\end{align} 
Here the subscript of $\langle r,x\rangle_2$ stands for the inner product 
over the second entry of the tensor. 
If the classical $r$-matrix is skew-symmetric $r_{12}=-r_{21}$\,,  
the CYBE \eqref{CYBE-ten} is equivalently written as 
\begin{align} 
[R(x),R(y)]-R([R(x),y]+[x,R(y)])=0 \qquad \text{for}\qquad {}^\forall x,y\in \alg{g} 
\label{CYBE}
\end{align}
in terms of the linear $R$-operator. 

An important generalization of the CYBE is the 
{\it modified classical Yang-Baxter equation (mCYBE)}, that is 
\begin{align}
[R(x),R(y)]-R([R(x),y]+[x,R(y)])=-c^2[x,y] \qquad \text{for}\qquad {}^\forall x,y\in \alg{g}\,.  
\label{mCYBE}
\end{align}
with a parameter $c\in \mathbb{C}$\,. 
The mCYBE is reduced to the CYBE by taking the $c\to0$ limit. 
When $\alg{g}$ is a real Lie algebra, the possible non-zero parameters are $c=1$ or $i$
up to rescaling of the $R$-operator; $R\to |c|R$\,. 
Depending on these values, the associated $R$-operators are classified into  
{\it split} ($c=1$) and {\it non-split} ($c=i$) type, respectively. 
The $r$-matrices discussed in \cite{DMV2,DMV3} are of non-split type.

\subsection{Solutions of the Classical Yang-Baxter equation} 

In the following, we will introduce some solutions of the (m)CYBE. 

In this subsection, we work with  $\alg{g}=\alg{gl}(N)$ generated by 
$E_{ij}$ with $i,j=1,\cdots,N$ which satisfy the relations; 
\begin{align}
[E_{ij},E_{kl}]=\de_{kj}E_{il}-\de_{il}E_{kj}\,.  
\end{align}
For $\alg{gl}(N)$\,, the bilinear form used in \eqref{linear-R} is simply given by 
the trace of the fundamental representations,   
\begin{align} 
\langle E_{ij}, E_{kl}\rangle = \Tr(e_{ij}e_{kl}) =\de_{kj}\de_{il}\,,  
\end{align} 
where $(e_{ij})_{kl}=\de_{ik}\de_{jl}$\,. 

\subsubsection{Solutions of mCYBE} 

Let us first look at a solution of the mCYBE.  

\paragraph{\rm \bf Drinfeld-Jimbo $r$-matrix}

A typical solution of the {\it modified} CYBE is the Drinfeld-Jimbo type solution, 
\begin{align}
r_{\rm DJ}=c\sum_{1\leq i<j\leq N} E_{ij}\wedge E_{ji}
\label{DJ-ten}
\end{align}
with $c\in \mathbb{C}\backslash \{0\}$ and the notation $A\wedge B=A\otimes B-B\otimes A$\,. 
The associated linear $R$-operator is 
\begin{align}
R_{\rm DJ}(E_{ij})=\begin{cases} +c E_{ij} & i<j \\ 0 & i=j \\ -c E_{ij} & i>j  \end{cases}\,. 
\end{align}
The $r$-matrix (\ref{DJ-ten}) with $c=i$ (non-split type) is used for a 
$q$-deformation of the AdS$_5\times$S$^5$ string action \cite{DMV2,DMV3}.  
More generally, the Drinfeld-Jimbo $r$-matrix is determined by fixing a borel  
subalgebra of $\alg{g}$\,. For instance, the other choices of $r$-matrices
for $\alg{psu}(2,2|4)$ are discussed in \cite{DMV3}.

\subsubsection{Solutions of CYBE} 

The next is to consider solutions of the CYBE \eqref{CYBE-ten}. 
For linear $R$-operators, see \eqref{CYBE}. 
As a remarkable feature, it is possible to restrict $r$-matrices to 
a subalgebra $\alg{gl}(M)\subset \alg{gl}(N)$ with $M< N$\,. 
For the $r$-matrix of Drinfeld-Jimbo type, such restrictions 
are not allowed  due to non-vanishing $c^2$-terms on the RHS in \eqref{mCYBE}. 

\paragraph{\rm \bf Trivial $r$-matrix} 
It is trivial that $r=0$ is a solution of the CYBE \eqref{CYBE-ten}
and it does not deform the original AdS$_5\times$S$^5$ string action, which will be 
introduced in next section. However, this fact is a very important   
feature of the CYBE-type deformations since it allows us to partially deform AdS$_5\times$S$^5$ background. 
Indeed, $R=0$ does {\it not} satisfy the mCYBE \eqref{mCYBE}. 
It seems quite difficult to discuss the partial deformations 
based on the $r$-matrices satisfying the mCYBE.  

\paragraph{\rm \bf Abelian $r$-matrix}

The simplest non-trivial examples are {\it abelian} $r$-matrices, 
which are composed of Cartan generators like  
\begin{align}
r_{\rm Abe}= E_{ii}\wedge E_{jj}  \qquad\text{with}\qquad i\neq j\,.  
\end{align}
Any linear combinations of them also satisfy the CYBE. 
In terms of the associated linear $R$-operator, it only acts on the Cartan generators;
\begin{align}
R_{\rm Abe}(E_{ii})=-E_{jj}\,, \qquad R_{\rm Abe}(E_{jj})=E_{ii}\,. 
\end{align}

\paragraph{\rm \bf Jordanian $r$-matrix}

Slightly non-trivial solutions are the so-called {\it Jordanian} $r$-matrices.   
They have non-zero Cartan charges and thus are nilpotent. 
The generic expression of them is 
\begin{align}
r_{\rm Jor}=E_{ij}\wedge (\al E_{ii}-\be E_{jj})-\ga \sum_{i<k<j} E_{ik}\wedge E_{kj}
\quad \text{with} \quad \begin{cases} \ga=\al+\be & \text{or} \\ \ga=0 \end{cases} 
\label{jor-para}
\end{align}
for $1\leq i<j\leq N$\,. 
The parameter relations are required to satisfy the CYBE. 
Eventually it contains an extra deformation parameter in addition to the overall scaling. 
The non-trivial action of the linear $R$-operator is written down, 
\begin{align}
R_{\rm Jor}(E_{ji})&=-\al E_{ii} +\be E_{jj}\,, &
R_{\rm Jor}(E_{jk})&= -\ga E_{ik}\,, \\
R_{\rm Jor}(E_{kk})&=(\al\de_{ki}-\be\de_{kj})E_{ij}\,,  &
R_{\rm Jor}(E_{ki})&= +\ga E_{kj} \,,
\end{align}
where $i<k<j$\,. Indeed, it is easy to see the nilpotency;
$(R_{\rm Jor})^n=0$ for $n\geq3$\,. 

When the parameters in \eqref{jor-para} take special values
$\al=\be=c$ and $\ga=2c$\,, the Jordanian $r$-matrix is obtained 
from the Drinfeld-Jimbo $r$-matrix \eqref{DJ-ten} through the 
{\it twisting} given by 
\begin{align}
[\Delta(E_{ij}),r_{\rm DJ}] = r_{\rm Jor}\qquad \text{with} \qquad 
\Delta(E_{ij})=E_{ij}\otimes 1 +1\otimes E_{ij}\,.  
\end{align}

\paragraph{\rm \bf Abelian-Jordanian $r$-matrix}

One may also consider $r$-matrices commuting each other and nilpotent. 
Such $r$-matrices can be constructed from commuting positive (or negative) generators and 
they may be referred as to {\it Abelian-Jordanian type}.  
A typical example takes the form 
\begin{align}
r_{\rm AJ}=E_{ij}\wedge E_{kl}\qquad \text{with} \qquad 
i<j\,, ~k<l\,,~ j\neq k\,,~ i\neq l \,. 
\end{align}
The associated linear $R$-operator is given by 
\begin{align}
R_{\rm AJ}(E_{lk})=E_{ij}\,, \qquad R_{\rm AJ}(E_{ji})=-E_{kl}\,, 
\end{align}
and acts trivially on the other generators. 
In this case, the nilpotency is $(R_{\rm AJ})^n=0$ for $n\geq2$\,.

\section{Deformed AdS$_5\times$S$^5$ string action  \label{Sec:action}} 

In the previous section, we have seen some typical solutions of the CYBE.  
Then, let us utilize them to construct deformed string actions on AdS$_5\times$S$^5$\,. 

\subsection{Definition of deformed string action}

A method to deform principal chiral models with classical 
$r$-matrices was introduced by Klimcik, which is called 
{\it the Yang-Baxter sigma models} \cite{Klimcik}. 
This technique has been generalized by Delduc-Magro-Vicedo 
not only for the bosonic cosets \cite{DMV} 
but also the AdS$_5\times$S$^5$ string action \cite{DMV2,DMV3}. 
The point is that these deformations are based on the Drinfeld-Jimbo
type $r$-matrix satisfying the mCYBE. 
In this section, we will give a short review of the models deformed by 
classical $r$-matrices satisfying the CYBE \cite{KMY-Jor}.

The deformed action of the AdS$_5\times$S$^5$ superstring is given by 
\begin{align}
S=-\frac{1}{4}(\ga^{\al\be}-\ep^{\al\be})\int^\infty_{-\infty}\!\!\!d\tau\int^{2\pi}_0\!\!\!d\sigma~
\str\left(A_\al d_+\frac{1}{1-\eta R_g\circ d_+} A_\be\right)
\label{action}
\end{align}
where the left-invariant one-form $A_\alpha$ is defined as 
\begin{align}
A_\al\equiv g^{-1}\partial_\alpha g\,, \qquad   g\in SU(2,2|4)\,. 
\label{def-current}
\end{align}
Here we work in the conformal gauge. $\ga^{\al\be}$ and $\ep^{\al\be}$ are the flat metric 
and the anti-symmetric tensor on the string world-sheet
normalized by $\ga^{\al\be}={\rm diag}(-1,+1)$ and $\ep^{\tau\sigma}=1$\,.   
The dressed operator $R_g$ is defined as 
\begin{align}
R_g(X)\equiv g^{-1}R(gXg^{-1})g\,.  
\label{Rg}
\end{align} 
Noting that both $R$- and $R_g$-operators satisfy the CYBE. 
The $R$-operator is related to the tensorial representation of a classical $r$-matrix through 
\begin{align}
&R(X)=\str_2[r(1\otimes X)]=\sum_i \bigl(a_i \str (b_iX)-b_i \str (a_iX)\bigr) 
\label{linearR} \\
&\text{with}\quad r=\sum_i a_i\wedge b_i\equiv \sum_i (a_i\otimes b_i-b_i\otimes a_i)\,. \el 
\end{align}
Since $a_i, b_i$ are $\alg{su}(2,2|4)$-valued, the supertrace $\str$ 
is needed for $(4|4)\times (4|4)$ supermatrix. 
The operators $d_\pm$ are given by the following,  
\begin{align}
d_\pm=\pm P_1+2P_2\mp P_3\,,  
\label{d}
\end{align}
where $P_i$ ($i=0,1,2,3$) are the projections to the $\mathbb{Z}_4$-graded 
components of $\alg{su}(2,2|4)$\,.  
$P_0\,,P_2$ and $P_1\,,P_3$ are the projectors to the bosonic and 
fermionic generators, respectively. 
In particular, $P_0(\alg{su}(2,2|4))$ is nothing but $\alg{so}(1,4)\oplus\alg{so}(5)$\,, 
which is a local symmetry of the action \eqref{action}.

It is noted that the kappa-symmetry of this model is also proved  
without specifying the form of classical $r$-matrix \cite{KMY-Jor}
and its derivation is similar to the $q$-deformed case \cite{DMV2}.

\medskip 

\subsection{The light-cone expression}

For later arguments, we will rewrite the action \eqref{action}
in terms of the light-cone coordinates; $x^\pm=(\tau\pm \sigma)/2$\,. 
Then, the light-cone currents are given by 
\begin{align}
A_\pm = A_\tau \pm A_\sigma\,. 
\end{align}
It is also convenient to introduce the deformed light-cone currents 
as follows; 
\begin{align}
J_\pm\equiv  \frac{1}{1\mp \eta R_g\circ d_\pm} A_\pm\,.  
\label{Jpm}
\end{align}
With these notations, the Lagrangian \eqref{action} is now simply written as 
\begin{align}
L=\tfrac{1}{4}\str\left(A_-d_+(J_+)\right)=\tfrac{1}{4}\str\left(A_+d_-(J_-)\right)\,. 
\label{action-lc}
\end{align}
It is also noted that, by solving the relation \eqref{Jpm} with respect to $A_\pm$\,,   
the Lagrangian has {\it dipole-like} expressions as follows;  
\begin{align}
L&=\tfrac{1}{4}\str\left[J_+d_-(J_-)\right] -\tfrac{\eta}{4} \str\left[d_+(J_+)R_g (d_-(J_-))\right]  \\
&=\tfrac{1}{4}\str\left[J_-d_+(J_+)\right] +\tfrac{\eta}{4} \str\left[d_+(J_+)R_g (d_-(J_-))\right] \,. 
\end{align}

\subsection{On-shell flat current} 

The equation of motion of Lagrangian \eqref{action-lc} is given by 
\begin{align}
{\cal E}\equiv \partial_-d_+(J_+)+\partial_+d_-(J_-) +[J_-,d_+(J_+)]+[J_+,d_-(J_-)]=0\,. 
\label{eom}
\end{align}
By the definition of the current $A_\pm$ in \eqref{def-current}\,, 
it satisfies the zero-curvature condition, 
\begin{align}
{\cal Z}\equiv \partial_+A_- - \partial_-A_+ +[A_+,A_-]=0\,.  
\label{zero-curv}
\end{align}
Rewriting ${\cal Z}$ in terms of the deformed current \eqref{Jpm}\,, 
we obtain the following expression, 
\begin{align}
{\cal Z}&=\partial_+J_- - \partial_-J_+ +[J_+,J_-] +\eta R_g({\cal E}) +\eta^2 {\rm CYBE}(d_+(J_+),d_-(J_-))\,, 
\end{align} 
where we have denoted the LHS of the CYBE \eqref{CYBE} with the dressed $R_g$-operator by 
\begin{align}
{\rm CYBE}(X,Y)\equiv [R_g(X),R_g(X)]-R_g\bigl([R_g(X),Y]+[X,R_g(Y)]\bigr)\,,  
\end{align}
which vanishes when $R$ satisfies the CYBE\footnote{%
If the $R$-operator satisfies the mCYBE rather than the CYBE, 
the condition \eqref{zero-curv} is modified as \cite{DMV2} 
\begin{align}
{\cal Z}=\partial_+J_- - \partial_-J_+ +[J_+,J_-] +\eta R_g({\cal E})+\eta^2[d_+(J_+),d_-(J_-)]\,.   
\end{align}
Thus, the deformed current $J_\mu$ is no longer flat even if the equation of motion is satisfied.  
}. %
This relation implies that the deformed current $J_\pm$ is also 
flat current if the equation of motion is satisfied ${\cal E}=0$\,. 
In this sense, $J_\pm$ is an {\it on-shell} flat current. 

\subsection{Lax pairs} 

We are ready to introduce the Lax pair for the deformed string action \eqref{action-lc}\,. 
They are given by\footnote{%
If the case of the Drinfeld-Jimbo type deformation, the Lax pair is given by \cite{DMV2} 
\begin{align}
\cL_\pm\equiv J_\pm^{(0)}+\la\sqrt{1+\eta^2} J_\pm^{(1)}
+\la^{\pm2}\frac{1+\eta^2}{1-\eta^2}J_\pm^{(2)}+\la^{-1}\sqrt{1+\eta^2}J_\pm^{(3)}\,.  
\end{align}
} %
\begin{align}
\cL_\pm\equiv J_\pm^{(0)}+\la J_\pm^{(1)}+\la^{\pm2}J_\pm^{(2)}+\la^{-1}J_\pm^{(3)}\,,  
\label{Lax}
\end{align}
with a spectral parameter $\la\in \mathbb{C}$\,, 
where the $\mathbb{Z}_4$-grading is denoted by 
$J_\pm^{(k)}\equiv P_k(J_\pm)$ for $k=0,1,2,3$\,.  
Then, the equation of motion \eqref{eom} and flatness condition \eqref{zero-curv} 
are equivalent with the zero-curvature condition of the Lax pair; 
\begin{align}
\partial_+\cL_- - \partial_-\cL_+ +[\cL_+,\cL_-] =0\,. 
\end{align} 
Thus, the deformed string action \eqref{action-lc} is classically integrable in the sense that 
the Lax pair exists.

\section{CYBE deformation as non-local gauge transformation \label{Sec:twist}} 

As a remarkable feature of the CYBE-type deformation, 
we will show that the deformed current $J_\pm$ in \eqref{Jpm} is obtained as a consequence 
of a non-local gauge transformation from the undeformed current $A_\pm=g^{-1}\partial_\pm g$\,. 
More precisely, the deformed current can be expressed as 
\begin{align}
J_\pm=\tilde g^{-1} \partial_\pm \tilde g\,,
\label{ren-J}
\end{align}
where a group element $\tilde g$ is related to the undeformed one $g$
by a {\it twist operator} $\cF$ like 
\begin{align}
\tilde g = \cF^{-1} g \,. 
\end{align}
As we will see here, such an operator $\cF$ does exist when 
the $R$-operator satisfies the CYBE. 

\subsection{Twist operator } 

To construct the operator $\cF$\,, let us consider the following gauge transformation of $J_\pm$\,: 
\begin{align}
J_\pm^g\equiv g J_\pm g^{-1}-\partial_\pm g g^{-1} 
\qquad \Longleftrightarrow \qquad 
\partial_\pm+ J_\pm^g =g (\partial_\pm+ J_\pm)g^{-1}\,. 
\label{GT1}
\end{align} 
The current $J_\pm$ is also flat when equation of motion 
is satisfied. Hence the resulting current $J_\pm^g$ is obviously flat 
and explicitly computed as 
\begin{align}
J_\pm^g=g (J_\pm -A_\pm )g^{-1} 
=g\left(\frac{\pm\eta R_g\circ d_\pm }{1\mp \eta R_g\circ d_\pm}A_\pm\right)g^{-1}
=\pm\eta R\left(gd_\pm(J_\pm)g^{-1}\right)\,. 
\end{align}

The flatness of $J_\pm^g$ can be seen as a consistency condition of auxiliary linear problem; 
\begin{align}
\partial_\tau \cF &= -J_\tau^g \cF\,, \el \\
\partial_\sigma \cF &= -J_\sigma^g \cF \,,   
\label{lin-pro}
\end{align}
where $\cF\equiv \cF(\sigma,\tau)$ is a $(4|4)\times (4|4)$ matrix. 
This problem is formally solved up to world-sheet boundary term $\cF(0,\tau)$ as follows; 
\begin{align}
\cF(\sigma,\tau) = \pexp \left[-\int_0^\sigma \!\! d\sigma \, J_\sigma^g \right] \cF(0,\tau)\,.  
\end{align}
The equations \eqref{lin-pro} allow us to express the current $J_\pm^g$ as 
\begin{align}
J_\pm^g=-\partial_\pm \cF \cF^{-1}\,. 
\end{align}
Combining the gauge transformation \eqref{GT1} and the above relation, 
we notice that the deformed current $J_\pm$ could be casted into the following nice form, 
\begin{align}
J_\pm =g^{-1} J_\pm^g g +A_\pm 
=-g^{-1} \partial_\pm \cF \cF^{-1} g +g^{-1}\partial_\pm g
=(\cF^{-1}g)^{-1}\partial_\pm(\cF^{-1}g) \,.  
\end{align}
This expression is nothing but the desired form \eqref{ren-J} with  
$\tilde g\equiv \cF^{-1}g$\,. 
Furthermore, by introducing the difference between $g$ and $\tilde g$ as 
\begin{align}
\cG\equiv \tilde{g}^{-1} g= g^{-1} \cF g \,, 
\end{align}
we notice that the deformed current $J_\pm$ is directly obtained
from $A_\pm$ by the following non-local gauge transformation; 
\begin{align}
J_\pm= \cG A_\pm \cG^{-1}-\partial_\pm \cG \cG^{-1} 
\qquad \Longleftrightarrow \qquad 
\partial_\pm+J_\pm=\cG (\partial_\pm+A_\pm)\cG^{-1}\,. 
\end{align}

\subsection{Conserved current} 

The conserved current for the deformed model is easily 
obtained by the equation of motion \eqref{eom}\,.   
To find out the conserved current, we introduce  
\begin{align}
\Lambda^\al\equiv 2\ga^{\al\be} J_\be^{(2)}+\ep^{\al\be}(J_\be^{(1)}-J_\be^{(3)})\,.  
\end{align}
Then, by using this quantity, the projected current is expressed as 
\begin{align}
d_\pm(J_\pm)=\Lambda^\tau\pm \Lambda^\sigma\,. 
\end{align}
The equation of motion \eqref{eom} is now written as 
\begin{align}
{\cal E}=-2\left(\partial_\al\Lambda^\al+[J_\al,\Lambda^\al] \right)=0\,. 
\end{align}
This relation leads us to define the following conserved current  
\begin{align}
k^\al\equiv \tilde g \Lambda^\al \tilde g^{-1}\,. 
\end{align}
In fact, the conservation law $\partial_\al k^{\al}=0$ immediately follows from 
the expression \eqref{ren-J}. 

Here it is worth giving some comments on the symmetry algebra of the conserved charges. 
Firstly, we should note that the quantities $\oint k^\al$ are not conserved 
in general due to the twisted boundary condition. 
Thus, to construct the conserved charges, it would be necessary to suppose the 
infinitely extended world-sheet with boundary conditions;
\begin{align}
g\to {\rm const.} \qquad \text{when} \qquad \sigma\to \pm \infty\,. 
\end{align}
With these asymptotic boundary conditions, the conserved charges can be defined as 
\begin{align}
Q^\al\equiv \int_{-\infty}^{\infty} \!\! d\sigma k^{\al} \,. 
\label{charge}
\end{align}
One may expect that the resulting symmetry algebra would be the undeformed $\alg{psu}(2,2|4)$\,.  
It is, however, quite sensitive to the choice of the constant matrix $K$ which appears in 
\begin{align}
\cF(\sigma,\tau) = \pexp \left[-\int_{-\infty}^\sigma \!\! d\sigma \, J_\sigma^g \right] K\,.  
\end{align}
Indeed, we have a degree of freedom to multiplying such a constant matrix in the linear 
problem \eqref{lin-pro}. This matrix corresponds to boundary conditions of the string world-sheet. 
The conserved current is transformed as 
\begin{align}
k^\al\to K k^\al K^{-1} \qquad \text{under} \qquad \cF\to\cF K \,. 
\end{align} 
Then, it triggers the mixing of the conserved charges defined in \eqref{charge}. 
In fact, in the case of three-dimensional Schr\"odinger spacetimes, 
a suitable choice of the $K$-matrix brings a $q$-deformed Poincar\'e 
algebra \cite{Ohn, KY-Sch} and the associated infinite dimensional algebra (called 
an {\it exotic symmetry} \cite{KY-exotic}) into a standard ${\cal U}(\alg{sl}(2))$ and 
a Yangian symmetry ${\cal Y}(\alg{sl}(2))$ \cite{KMY-Jor3d}.

\section{From CYBE to deformations of AdS$_5\times$S$^5$ \label{Sec:CYBEtograv}}

The deformed action \eqref{action} is rather algebraic 
and hence it is not so obvious to deduce the background metric.  
We demonstrate here how to read off the metric
and the NS-NS two-form from the deformed action. 
Then we present some examples of the gravity/CYBE correspondence. 

To read off the metric and NS-NS two-form, it is sufficient to see the bosonic part of the deformed action. 
Firstly, we need to introduce the coordinate system $x^i$ of AdS$_5\times$S$^5$ via a parameterization 
of the group element;  
\begin{align}
g=\exp\bigl(\textstyle{\sum_i} x^i T_i\bigr) \quad \in \quad SU(2,2)\otimes SU(4)\,,  
\end{align}
where $T^i$ belong to the fundamental representation of $\alg{su}(2,2)\oplus\alg{su}(4)$\,. 
The coordinate system $x^i$ may be the Poincar\'e coordinates or 
the global coordinates, depending on the purpose.  
In the practical computation, it is often useful to utilize gamma-matrices as the generators $T_i$\,. 

Secondary, plugging the projector \eqref{d} with \eqref{action}, 
the bosonic part of the Lagrangian reads
\begin{align}
L=-\frac{1}{2}(\ga^{\al\be}-\ep^{\al\be}) \Tr[A_\al P_2(J_\be)] 
\qquad \text{with} \qquad 
J_\be=\frac{1}{1-2\eta R_g\circ P_2}A_\be \,. 
\label{bosL}
\end{align}
The metric and NS-NS two-forms are obtained as the symmetric and 
skew-symmetric part with respect to the world-sheet 
coordinates in the above expression, respectively.  
Then, the remaining is to evaluate the projected deformed bosonic current $P_2(J_\be)$
in \eqref{bosL} explicitly. 
This is easily done by multiplying $P_2\circ (1-2\eta R_g\circ P_2)$ on $J_\be$ and 
the following form is obtained, 
\begin{align}
(1-2\eta P_2\circ R_g )P_2(J_\be) = P_2(A_\be) \,. 
\end{align}
The above relation may be regarded  as the definition of $P_2(J_\be)$\,. 
This relation takes the value on 
$\alg{su}(2,2)\oplus\alg{su}(4)$\,, but the 
projector $P_2$ reduces the number of the linearly independent relations 
just to $10=5+5=\dim$(AdS$_5\times$S$^5$)\,.   
In particular, when a deformation of either AdS$_5$ or 
S$^5$ is considered, only five linear relations are concerned with the analysis. 

\subsection{Examples \label{Subsec:exam}} 

We will present some examples of deformed backgrounds, which are obtained from 
typical solutions of the CYBE (explained in Section \ref{Sec:classical-R})\footnote{
We set that $\eta=1$ in \eqref{bosL} here.}. 

\begin{table}[t]
\caption{\label{table} A catalog of the gravity/CYBE correspondence. 
Here we list some typical examples of classical $r$-matrices and 
the associated gravity backgrounds, which are 
explained in the Subsection \ref{Subsec:exam} and Section \ref{Sec:T11}. } 
\vspace*{0.2cm}
\begin{center}
\begin{tabular}{lcl}
\br
Classical $r$-matrices & $\to$\quad  & Deformed gravity solutions \\
\mr
Trivial: $r=0$ && AdS$_5\times$S$^5$ \cite{MT} \\
Jordanian: $r_{\rm Jor}=E_{24}\wedge (E_{22}-E_{44})$ & & Melvin twists \cite{HRR} \\ 
Abelian: $r_{\rm Abe}=h_1\wedge h_2$ with $h_i\in \alg{su}(4)$ & & $\gamma_i$-deformations of S$^5$ \cite{LM,Frolov} \\ 
Abelian: $r_{\rm Abe}=h_1\wedge h_2$ with $h_i\in \alg{su}(2,2)$& & $\gamma_i$-deformations of AdS$_5$ 
\cite{MS,DHH,STV,Kul} \\ 
Abelian Jordanian: $r_{\rm AJ} = p_1\wedge p_2$ & & Gravity duals for NC Yang-Mills \cite{HI,MR} \\ 
\mr
Abelian: $r_{\rm Abe}=K_3\wedge L_3$ & & $\gamma_i$-deformations of $T^{1,1}$ \cite{LM,CO} \\ 
\br
\end{tabular}
\end{center}
\end{table}

\subsubsection{Three parameter $\hat\ga_i$-deformation of S$^5$} 

Firstly, let us consider an abelian classical $r$-matrix with $\alg{su}(4)$\,, 
which consists of the three Cartan generators $h_1,h_2,h_3$\,;  
\begin{align}
r_{\rm Abe}=\frac{1}{8}(\hat\ga_3 h_1\wedge h_2+\hat\ga_1 h_2\wedge h_3+\hat\ga_2 h_3\wedge h_1 )\,,  
\label{abe-S5}
\end{align}
where $\hat\ga_i$ ($i=1,2,3$) are deformation parameters.  
This $r$-matrix deforms only the S$^5$ part. 

The resulting metric and NS-NS two-forms turn out to be \cite{MY-LM}
\begin{align}
ds^2 &= ds^2_{\rm AdS_5} + \sum_{i=1}^3(d\rho_i^2+G \rho_i^2d\phi_i^2) 
+ G \rho_1^2\rho_2^2\rho_3^2 \Bigl(\sum_{i=1}^3\hat{\ga}_i d\phi_i\Bigr)^2  \,, 
\\ 
B_2 &= G \left(
\hat{\ga}_3\rho_1^2\rho_2^2\,d\phi_1\wedge d\phi_2 + \hat{\ga}_1\rho_2^2\rho_3^2\,d\phi_2\wedge d\phi_3  
+ \hat{\ga}_2\rho_3^2\rho_1^2\,d\phi_3\wedge d\phi_1  \right) \,, 
\end{align}
where $\rho_i$ satisfy the following constraints and the scalar function 
$G$ is defined by 
\begin{align}
\rho_1^2+\rho_2^2+\rho_3^2 =1\,,  \qquad 
G^{-1} \equiv 1 + \hat{\ga}_3^2\rho_1^2\rho_2^2 + \hat{\ga}_1^2\rho_2^2\rho_3^2 + \hat{\ga}_2^2\rho_3^2\rho_1^2\,. 
\end{align} 
These backgrounds agree with the well-known three-parameter $\hat\ga_i$-deformations of S$^5$ \cite{LM,Frolov}. 
In particular, it reduces to the Lunin-Maldacena case \cite{LM} when $\hat\ga_1=\hat\ga_2=\hat\ga_3$\,. 
As pointed out in \cite{Frolov}, these deformed backgrounds are reproduced by performing TsT transformations 
(which is a sequence of two T-dualities and a shift).

\subsubsection{Gravity duals for noncommutative gauge theories} 

The next example is an abelian Jordanian classical $r$-matrix with $\alg{su}(2,2)=\alg{so}(2,4)$\,; 
\begin{align}
r_{\rm AJ}=\frac{1}{2}\bigl(a^2 p_2\wedge p_3+a'^2 p_0\wedge p_1 \bigr)\,.  
\end{align}
Here $p_\mu$ are translation generators of the conformal algebra 
and commute each other $[p_\mu,p_\nu]=0$\,.  
Thus, it trivially satisfies the CYBE. 
The explicit matrix representations of $p_\mu$ are given in \cite{MY-MR}.  
It is noted that this $r$-matrix only deforms the AdS$_5$ part. 

The resulting metric and NS-NS two-form in the Poincar\'e coordinates are \cite{MY-MR} 
\begin{align}
ds^2 &= \frac{z^2 }{z^4-4a'^4}(-(dx^0)^2 + (dx^1)^2)+\frac{z^2 }{z^4+4a^4}((dx^2)^2 + (dx^3)^2)
+\frac{dz^2}{z^2}+ds_{{\rm S}^5}^2\,,   \\
B_2&=-\frac{a'^2}{z^4 - a'^4 }dx^0\wedge dx^1 +\frac{a^2}{z^4 + a^4}dx^2\wedge dx^3\,.   
\end{align}
These completely agree with the gravity duals for non-commutative gauge theories \cite{HI, MR}. 

It is worth mentioning that the classical integrability of this background
automatically follows from our formulation as a byproduct. It is because the Lax pair has already been 
constructed in a generic form \eqref{Lax}.

One may also consider abelian twists of AdS$_5$ rather than S$^5$\,. 
For this purpose, one can simply adopt the abelian $r$-matrix consisting of 
the Cartan generators in $\alg{su}(2,2)$ instead of \eqref{abe-S5}. 
The resulting metric and NS-NS two-form agree with \cite{MS} and also 
include \cite{DHH} as a particular case, which is obtained by means of a Melvin twist.  

\subsubsection{New backgrounds from Jordanian $r$-matrix}  

We can also consider new integrable backgrounds which are obtained by 
Jordanian $r$-matrices such as \eqref{jor-para}.  
For instance, it is easy to see that the following $r$-matrix satisfies the CYBE,  
\begin{align}
r_{\rm Jor}=E_{24}\wedge (\ga E_{22}-\ga^* E_{44}) 
\qquad \text{with} \qquad \ga\in \mathbb{C}\,,   
\label{jor-r}
\end{align}
where $(E_{ij})_{kl}=\de_{ik}\de_{jl}$ are the fundamental representations of $\alg{su}(2,2)$
and $\ga^*$ denotes the complex conjugate of $\ga\in \mathbb{C}$\,.  

The resulting metric of AdS$_5$ and NS-NS two-form turn out to be \cite{MY-TsT,KMY-Jor}\,,  
\begin{align}
ds^2&=\frac{-2dx^+dx^- +(dx^1)^2+(dx^2)^2+dz^2}{2 z^2}
-\frac{|\ga|^2 \bigl((x^1)^2 +(x^2)^2\bigr)+({\rm Re}\ga)^2z^2 }{2 z^6}(dx^+)^2\,, \el \\
B_2&=-{\rm Re}\ga \frac{ \bigl(x^1dx^1+x^2dx^2+zdz\bigr)\wedge dx^+}{2z^4} 
+{\rm Im}\ga\frac{\bigl(x^2 dx^1-x^1 dx^2\bigr)\wedge dx^+}{2z^4}\,.     
\label{Jor-bg}
\end{align}
It is noted that both the metric and the NS-NS two-form are real even though 
the Jordanian $r$-matrix itself \eqref{jor-r} does depend on a complex deformation parameter $\ga$\,. 
When the parameter $\ga$ is real, the deformed background was computed in \cite{KMY-Jor} 
and the complete type IIB supergravity solutions are constructed in \cite{KMY-SUGRA}. 
In the case that $\ga$ is a pure imaginary, the above backgrounds reduce to 
the one obtained by means of a {\it null Melvin twist} \cite{HRR}.  
Furthermore, for a generic complex parameter $\ga$\,, 
the deformed backgrounds \eqref{Jor-bg} are obtained from the undeformed 
AdS$_5\times$S$^5$ by a chain of T-dualities and S-duality \cite{MY-TsT}.  
This strongly suggests that the Yang-Baxter sigma model with the Jordanian 
$r$-matrix \eqref{jor-r} gives rise to a consistent string background.

\section{An application to a non-integrable background \label{Sec:T11}}  

So far, we have concentrated on deformations of the AdS$_5\times$S$^5$ case. 
The AdS$_5\times$S$^5$ background is known as an integrable background 
because the metric can be represented by a supercoset with the $\mathbb{Z}_4$-grading \cite{BPR}. 
An interesting question is whether or not our deformation technique is applicable to 
non-integrable backgrounds. In the recent progress, 
many examples of non-integrable backgrounds have been discovered.  
Among them, we will consider type IIB string theory on the AdS$_5\times T^{1,1}$ background. 
This theory is known to be dual for the ${\cal N}=1$ superconformal field theory 
in four dimensions \cite{KW}. 
The non-integrability of this background is shown in \cite{BZ} by presenting the existence of chaotic string solutions. 
The next issue that we would like to consider is deformations of this background 
by utilizing the Yang-Baxter sigma model approach. 
Interestingly, it turns out that it successfully works even for this non-integrable background \cite{CMY-T11}. 
The point is that, to apply the Yang-Baxter sigma model description,  
the parent background needs to be represented by a coset, although it is not necessary to be symmetric.

\subsection{A coset construction of $T^{1,1}$}

Let us consider the $T^{1,1}$ part of the AdS$_5\times T^{1,1}$ background. 
The first step to consider CYBE-type deformations of $T^{1,1}$ is to find out a coset construction. 
A five-dimensional Sasaki-Einstein manifold $T^{1,1}$ is represented by a $U(1)$-fibration
over $S^2\times S^2$ and it has a coset structure like $[SU(2)\times SU(2)]/U(1)$ as denoted in 
\cite{CD, Gauntlett:2004yd}. 
This coset description, however, does not leads the correct metric of $T^{1,1}$
without suitable rescalings of vierbeins \cite{R,Castellani:1983tb}. 
For our purpose, we would like to avoid the rescaling because the dressed $R$-operator
\eqref{Rg} depends on a group element $g$ itself. 
In order to overcome this difficulty,  we found another coset description of $T^{1,1}$\,. 
With this coset, it is not necessary to perform any rescaling so as to obtain the $T^{1,1}$ metric. 
The coset description we found in \cite{CMY-T11} is given by  
\begin{align}
T^{1,1}=\frac{SU(2)\times SU(2)\times U(1)_R}{U(1)_1\times U(1)_2}\,.   
\label{T11-coset}
\end{align}
With this description, the manifold obviously has the $SU(2)\times SU(2)\times U(1)_R$
symmetry, where $SU(2)\times SU(2)$ and 
$U(1)_R$ correspond to a flavor symmetry and 
an R-symmetry in the dual gauge theory, respectively. 
Let $T_i$ ($i=1,2$) be the generators of two $U(1)_i$'s in the denominator. 
Then, the two $U(1)_i$'s are embedded into the numerator as 
\begin{align}
T_1=K_3+L_3\,, \qquad  T_2=K_3-L_3+4M \,, 
\end{align} 
where $K_i$ and $L_i$ ($i=1,2,3$) are the generators of the two $\alg{su}(2)$'s, respectively. 
In particular, $K_3$ and $L_3$ are the Cartan generators, and $M$ is the $\alg{u}(1)_R$ generator.  

By introducing a parameterization of the group element $g\in SU(2)\times SU(2)\times U(1)_R$
and defining the left-invariant one-form $A=g^{-1}dg$\,,   
the metric of $T^{1,1}$ can be reproduced as  
\cite{CMY-T11} 
\begin{align}
ds^2_{T^{1,1}} &=-\frac{1}{3}\str[AP(A)] \nonumber \\ 
&=\frac{1}{6}\sum_{i=1,2}(d\theta_i^2+\sin^2\theta_id\phi_i^2)
+\frac{1}{9}(d\psi+\cos\theta_1d\phi_1+\cos\theta_2d\phi_2)^2\,,   
\label{T11metric}
\end{align}
where the coset projector $P$ is defined by 
\begin{align}
P(A)& \equiv A-\frac{\str[T_1A]}{\str[T_1T_1]}T_1 - \frac{\str[T_2A]}{\str[T_2T_2]} T_2\,.    
\label{proj}
\end{align} 

One may wonder why the supertrace $\str$ is adopted here rather than the usual trace. 
Indeed, the superstrace in \eqref{T11metric} is assumed to be taken over the 
$(4|1)\times (4|1)$ supermatrices, which is the fundamental representation of $SU(2)\times SU(2)\times U(1)_R$\,. 
This may be justified by interpreting the matrix as a part of a bigger 
$(8|1)\times (8|1)$ supermatrix including an ${\cal N}=1$ superconformal algebra as follows; 
\begin{align}
\left(
\begin{array}{cc|c} 
SU(2) & 0 & 0 \\ 
 0 & SU(2) & 0  \\ \hline  
 0 & 0 & U(1)_R  
\end{array}
\right)
\quad \hookrightarrow  \quad 
\left(
\begin{array}{ccc|c} 
SU(2,2) & 0 & 0 & \overline{F}{}^A \\ 
0 & SU(2) & 0 & 0 \\ 
0 & 0 & SU(2) & 0  \\ \hline  
F_A & 0 & 0 & U(1)_R  
\end{array}
\right)\,.  
\end{align}
The lines in the matrices distinguish the $\mathbb{Z}_2$-grading.  
In the right big supermatrix, an ${\cal N}=1$ superconformal algebra
$PSU(2,2|1)$ sitting at the four corners together with the supercharges 
$F_A, \overline{F}{}^A$ ($A=1,\cdots,4$).  
The point is that the $U(1)_R$ symmetry is originated in 
the superconformal symmetry $PSU(2,2|1)$\,.  
In other words, the coset \eqref{T11-coset} should be regarded as a subsector of the supercoset; 
\begin{align}
\frac{PSU(2,2|1)\times SU(2)\times SU(2)}{SO(1,4)\times U(1)_1\times U(1)_2}\,.   
\end{align}

\subsection{Deformations of $T^{1,1}$ as Yang-Baxter sigma models}

We have already revealed the coset structure of $T^{1,1}$\,. We are now ready to apply the 
deformation technique based on classical $r$-matrices. 
The Lagrangian of the Yang-Baxter sigma model on $T^{1,1}$ is given by 
\begin{align}
L \equiv \frac{1}{3}(\ga^{\al\be}-\ep^{\al\be})\str \left[A_\al P\frac{1}{1-2R_g\circ P} A_\be\right] \,, 
\end{align}
with the projector $P$ defined in \eqref{proj}.  

As a natural generalization of the S$^5$ case \eqref{abe-S5}, 
let us consider an abelian $r$-matrix which is composed of the three Cartan generators
of $\alg{su}(2)\oplus\alg{su}(2)\oplus\alg{u}(1)_R$\,,    
\begin{align}
r_{\rm Abe} =\frac{1}{3} \hat\ga_1 L_3\wedge M+\frac{1}{3} \hat\ga_2  M\wedge K_3 
-\frac{1}{6} \hat\ga_3 K_3\wedge L_3\,, 
\label{abe-3para}
\end{align}
with deformation parameters $\hat\ga_i$ ($i=1,2,3$). 

After some computations, the resulting metric and NS-NS two-from 
turn out to be \cite{CMY-T11} 
\begin{align}
ds^2 &= G(\hat\ga_1,\hat\ga_2,\hat\ga_3)\Bigl[
\frac{1}{6}\sum_{i=1,2}(G(\hat\ga_1,\hat\ga_2,\hat\ga_3)^{-1}d\theta_i^2+\sin^2\theta_i d\phi_i^2) \el \\
&\quad 
+\frac{1}{9}(d\psi+\cos\theta_1d\phi_1+\cos\theta_2d\phi_2)^2 
+\frac{\sin^2\theta_1\sin^2\theta_2}{324}(\hat\ga_3 d\psi+\hat\ga_1 d\phi_1+\hat\ga_2 d\phi_2)^2 \Bigr]\,, 
\\
B_2 &=G(\hat\ga_1,\hat\ga_2,\hat\ga_3)\Bigl[
\Bigl\{\hat\ga_3 \Bigl(\frac{\sin^2\theta_1\sin^2\theta_2}{36}
+\frac{\cos^2\theta_1\sin^2\theta_2+\cos^2\theta_2\sin^2\theta_1}{54}\Bigr)\el \\
&\hspace{33mm}
-\hat\ga_2 \frac{\cos\theta_2\sin^2\theta_1}{54}
-\hat\ga_1 \frac{\cos\theta_1\sin^2\theta_2}{54} 
\Bigr\}d\phi_1\wedge d\phi_2 \el\\
&\quad 
+\frac{(\hat\ga_3\cos\theta_2-\hat\ga_2)\sin^2\theta_1}{54}d\phi_1\wedge d\psi 
-\frac{(\hat\ga_3\cos\theta_1-\hat\ga_1)\sin^2\theta_2}{54}d\phi_2\wedge d\psi 
\Bigr]\,, 
\end{align}
where the scalar function is defined as  
\begin{align}
G(\hat\ga_1,\hat\ga_2,\hat\ga_3)^{-1}&\equiv1
+\hat\ga_3^2 \Bigl(\frac{\sin^2\theta_1\sin^2\theta_2}{36}
+\frac{\cos^2\theta_1\sin^2\theta_2+\cos^2\theta_2\sin^2\theta_1}{54}\Bigr)
+\hat\ga_2^2 \frac{\sin^2\theta_1}{54} \el \\
&\qquad 
+\hat\ga_1^2 \frac{\sin^2\theta_2}{54} 
-\hat\ga_2\hat\ga_3 \frac{\sin^2\theta_1\cos\theta_2}{27} 
-\hat\ga_3\hat\ga_1 \frac{\sin^2\theta_2\cos\theta_1}{27} \,. 
\end{align}
Indeed, these expressions agree with the ones obtained in \cite{CO} via TsT-transformations.  
In particular, when $\hat\ga_1=\hat\ga_2=0$ and $\hat\ga_3$ is non-zero, 
the above deformed backgrounds reduce to the Lunin-Maldacena case \cite{LM}. 

\section{Conclusion and discussion \label{Sec:concl}}

In this article, we have given a short review of the recent progresses on 
novel relations between classical $r$-matrices satisfying the CYBE 
and deformed string backgrounds \cite{KMY-Jor,KMY-SUGRA,MY-LM,MY-MR,CMY-T11}. 
The relations are worthy of being called the {\it gravity/CYBE correspondence} \cite{MY-LM,MY-MR}. 
According to them, the solution space of the CYBE may be regarded as a moduli space of 
a certain class of deformed gravitational backgrounds. 
In particular, we have presented several important examples including 
three parameter $\hat\ga_i$-deformations of the S$^5$ part \cite{LM,Frolov} and 
the gravity duals of non-commutative ${\cal N}=4$ super Yang-Mills 
theories \cite{HI,MR} in Section \ref{Sec:CYBEtograv}. 
We have also discussed an application of the deformation technique 
to a non-integrable background, AdS$_5\times T^{1,1}$ \cite{CMY-T11} in Section \ref{Sec:T11}.  

There remain, however, several open questions to be answered. 
Let us list some of them in the following;  
\begin{itemize}

\item Firstly, we are not sure whether or not any deformed backgrounds obtained by classical $r$-matrices
are consistent with the equations of motion of type IIB supergravity. 
Even though we proved that the deformed models have kappa-symmetry in general \cite{KMY-Jor}  
and some important deformed supergravity backgrounds are successfully obtained 
as demonstrated in Section \ref{Sec:CYBEtograv},  
it is not straightforward to conclude that the resulting backgrounds always satisfy the equations of motion 
of type IIB supergravity. 

\item It is not obvious that mutually different $r$-matrices  
label pair-wise different backgrounds. 
It is observed in \cite{KMY-SUGRA} that two non-equivalent $r$-matrices lead to 
the identical metric, up to coordinate transformations. Thus, to establish the one to one correspondence
between $r$-matrices and deformed backgrounds, it would be necessary to 
take into account an appropriate quotient by algebra automorphisms. 

\item Can we find the corresponding $r$-matrix for a given deformed background? 
More generally, does there alway exist an $r$-matrix for any analytically deformed AdS$_5\times$S$^5$ background? 
Even though we know how to compute the deformed back metric and the NS-NS two-forms from the $r$-matrices, 
the inverse direction should also be clarified. 

\item Finally, it is desirable to figure out the composite rule of the $r$-matrices. 
On the gravity theory side, some deformed backgrounds are obtained by string dualities such as 
TsT-transformations and we could think of a chain of them to construct more complicated backgrounds. 
In this sense, we have a structure of product of the dualities. 
The question is how it works for the classical $r$-matrices if the gravity/CYBE correspondence holds.   
We expect that the twist operator discussed in Section \ref{Sec:twist} might be a clue to solve
this problem.
 
\end{itemize}

Even though there are still many issues to establish the gravity/CYBE correspondence, 
we believe that classical $r$-matrices are very fascinating objects to characterize 
deformed gravity solutions. As a final remark, the deformation technique presented here preserves 
the classical integrability of the undeformed background, 
but it can describe a non-integrable deformation of a non-integrable background, AdS$_5\times T^{1,1}$ 
\cite{CMY-T11}. This result indicates that the gravity/CYBE correspondence is not restricted to 
a class of integrable backgrounds and capture a wider class of gravity solutions. 
It is an important task to ascertain the limitation of this scenario. 
We hope to report on some new results along this direction in the near future.

\section*{Acknowledgment}

We are grateful to Gleb Arutyunov, Riccardo Borsato, Takashi Kameyama  
and Stefan Vandoren for valuable comments and discussions. 
We also thank Io Kawaguchi for his collaboration at the beginning of this project. 
Especially, we would like to appreciate P. Marcos Crichigno for the collaboration 
concerning with \cite{CMY-T11} and comments on this manuscript. 
T.M.\ is supported by the Netherlands Organization for Scientific 
Research (NWO) under the VICI grant 680-47-602. 
This work is also part of the ERC Advanced grant research programme 
No.~246974, ``Supersymmetry: a window to non-perturbative physics" 
and of the D-ITP consortium, a program of the NWO that is funded by the 
Dutch Ministry of Education, Culture and Science (OCW).

\end{document}